# FAULT TOLERANT REAL TIME SYSTEMS

A. Christy Persya,Sr.Lecturer,

Department of Information Science and Engineering,

The Oxford College of Engineering,

Bangalore., India

christypersya@gmail.com

T.R.Gopalakrishnan Nair Ph.D.

Director, Research and Industry

DS Institutions

Bangalore, India

trgnair@ieee.org

**ABSTRACT**

*Real time systems are systems in which there is a commitment for timely response by the computer to external stimuli. Real time applications have to function correctly even in presence of faults. Fault tolerance can be achieved by either hardware or software or time redundancy. Safety-critical applications have strict time and cost constraints, which means that not only faults have to be tolerated but also the constraints should be satisfied. Deadline scheduling means that the taskwith the earliest required response time is processed. The most common scheduling algorithms are :Rate Monotonic(RM) and Earliest deadline first(EDF).This paper deals with the interaction between the fault tolerant strategy and the EDF real time scheduling strategy.*

## 1. Introduction

Real-time systems can be classified as hard real time systems in which the consequences of missing a deadline can be catastrophic and soft real time systems in which the consequences are relatively tolerable. In hard real time systems it is important that tasks complete within their deadline even in the presence of a failure. Examples of hard real-time systems are control systems in space stations, auto pilot systems and monitoring systems for patients with critical conditions. In soft real-time systems it is more important to economically detect a fault as soon as possible rather than to mask a fault. Examples of soft real-time systems are all kind of airline reservation, banking, and E-commerce applications. The following sections reviews the fault tolerant strategy and EDF scheduler strategy. Section 2 summaries fault-tolerant techniques and section 3 discuss the Fault Tolerant Deadline scheduling strategy and section 5 concludes the paper.

## 2. Techniques for Fault Tolerance

Fault tolerance is the ability to continue operating despite the failure of a limited subset of their hardware or software. So the goal of the system designer is to ensure that the probability of system failure is acceptably small. There can be either hardware fault or software fault, which disturbs the real time systems to meet their deadlines.

### 2.1 Fault Types

There are three types of faults: Permanent, intermittent, and transient. A permanent fault does not die away with time, but remains until it is repaired as the affected unit is replaced. This is an intermittent fault cycle between the fault–active and fault benign states. A transient fault dies away after some time.

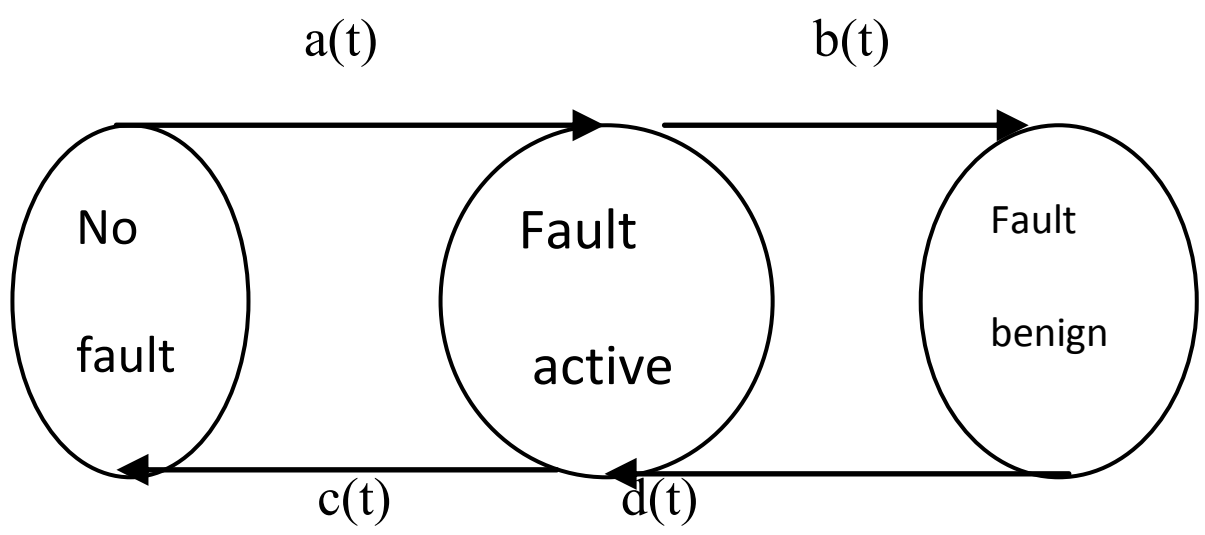

Fig 2.1 State Diagram of the Fault Classes

| Fault Type | Condition |
|---|---|
| Permanent | a(t) > 0 , b(t) = c(t) = d(t) = 0 |
| Transient | a(t) > 0 , b(t) = 0 , c(t) > 0 , d(t) = 0 |
| Intermittent | a(t) > 0 , b(t) > 0 , c(t) = 0 , d(t) > 0 |

In the figure 2.1, a(t) and b(t) are the rates at which the fault switches states, and t is the age of the fault.

### 2.2 Fault Detection

Fault detection can be done either online or offline. Online detection goes on in parallel with normal system operation. Offline detection consists of running diagnostic tests.

#### 2.2.1 Error Detection Techniques

In order to achieve fault tolerance, the first requirement is that transient faults have to be detected. Several error-detection techniques are there against transient faults: watchdogs, duplication and few others.

**Watchdogs.** In the case of watchdogs[2] program flow or transmitted data is periodically checked for the presence of errors. The simplest watchdog schema, *watchdog timer*, monitors the execution time of processes, whether it exceeds a certain limit.

**Duplication.** Duplication is an approach to have multiple processors, which are supposed to put out the same result and compare the results. A discrepancy indicates the existence of a fault .

There are several other error-detections techniques, e.g. signatures, assertions or the widely-used parity-bit check.

#### 2.2.2 Redundancy

Fault tolerance system is to be kept running despite the failure of some of its parts, it must have spare capacity to begin.
There are two ways to make a system more resistant to faults[3].
-Hardware: this technique relies on adding extra redundant hardware to a system to make it fault-tolerant.
-Software: this technique relies on duplicating the code, process, or even messages, depending on the context.
A typical example of where the above techniques are applied would be the autopilot system on-board a large-sized passenger aircraft[4].
A passenger aircraft typically consists of a central autopilot system with two other backups. This is an example of making a system with two other backups. This is an example of making a system fault tolerant by adding redundant hardware. The two extra systems will not be used unless the main system is completely broken.
However, this is not sufficient, since in the event that the main system starts behaving erratically the lives of many people is in danger. The system is therefore also made resistant to faults using software.
Generally, every process of the autopilot runs more than two copies, distributed across different computers. The system then votes on the results of these process. To make the system even more secure, some autopilots also employ the principle of design diversity. In this feature, not only a software is run multiple times, but also each copy is written by a different engineering team. The likelihood of same mistake being made by different engineering teams is very low.
However, such measures are only applied for highly critical systems. In general, hardware redundancy is avoided as far as possible, due to limited resources that are available. Weight of the system, power consumption, and price constraints make it difficult to employ high hardware redundancy to make the system fault tolerant. Software redundancy is therefore, more commonly used to increase fault tolerance of systems.
There are few factors that affect the diversity of the multiple versions. The first factor is the requirements specification. A mistake in the specification causes a wrong output to be delivered.
A second approach is the programming language. The nature of the language affects the programming style greatly.
A third factor is the numerical algorithms that are used. Algorithms implemented to a finite precision can behave quite differently for certain sets of inputs than do theoretical algorithms, which assume infinite precision.
A fourth factor is the nature of the tools that are being used, the probability of common-mode failure might increase. A fifth factor is the training and quality of the programmers and the management structure. The major difficulty in software is labor-intensive.

### 2.3 Fault Tolerance Techniques

1*) TMR (Triple Modular Redundancy)*
Multiple copies are executed and error checking is achieved by comparing results after completion. In this scheme, the overhead is always on the order of the number of copies running simultaneously.
*2) PB (Primary/Backup)*
The tasks are assumed to be periodic and two instances of each task (a primary and a backup) are scheduled on a uni-processor system. One of the restrictions of this approach is that the period of any task should be a multiple of the period of its preceding tasks. It also assumes that the execution time of the backup is shorter than that of the primary.
*3) PE (Primary/Exception)*

It is the same as PB method except that exception handlers are executed instead of backup programs.

***Primary Backup Fault Tolerance***

This is the traditional fault-tolerant approach wherein both time as well as space exclusions are used. The main idea behind this algorithm is that (a) the backup of a task need not execute if its primary executes successfully, (b) the time exclusion in this algorithm ensures that no resource conflicts occur between the two versions of any task, which might improve the schedulability. Disadvantages in this system are that (a) there is no de-allocation of the backup copy, (b) the algorithm assumes that the tasks are periodic (the times of the tasks are predetermined), (c) compatible (the period of one process is an integral multiple of the period of the other process) and execution time of the backup is shorter than that of the primary process.

## 3 Fault tolerant Deadline Scheduling

*A. Backup Overloading Scheduling Algorithm*
The following steps form the procedure used to implement the backup overloading algorithm.

*1) Arriving task*
A task has four properties when it arrives, arrival time (ai), Ready time (ri), Deadline – (di) and worst case computation time (ci) represented as Ti = (ai, ri, di, ci)

*2) EDF schedulability*
Check if all the tasks can be scheduled successfully using the earliest deadline first algorithm. If the schedulability test fails, then reject the set of tasks saying that they are not schedulable.

*3) Searching for timeslot*
When task Ti arrives, check each processor to find if the primary copy (Pri) of the task can be scheduled between ri and di. Say it is scheduled on processor Pi.

*4) Try overloading*
Try to overload the backup copy (Bki) on an existing backup slot on any processor other than Pi.
Note: The backups of 2 primary tasks that are scheduled on the same processor must not overlap. If the processor fails, it will not be possible to schedule the two backups simultaneously since they are on the same time slot (overloaded).

*5) EDF Algorithm*
If there is no existing backup slot that can be overloaded, then schedule the backup on the latest possible free slot depending upon the dead line of the task. The task with the earliest deadline is scheduled first.

*6) De-Allocation of backups*
If a schedule has been found for both the primary and backup copy for a task, commit the task, otherwise reject it. If the primary copy executes successfully, the corresponding backup copy is de-allocated.

*7) Backup execution*
If there is a permanent or transient fault in the processor, the processor crashes and then all the backups of the tasks that were running on this system are executed on different processors.

*B. A feasible overloading example*
Figure 3.1 shows 4 processes running on 3 different processors P1, P2 and P3. The backup copies of these processes are scheduled to run on the different processors.

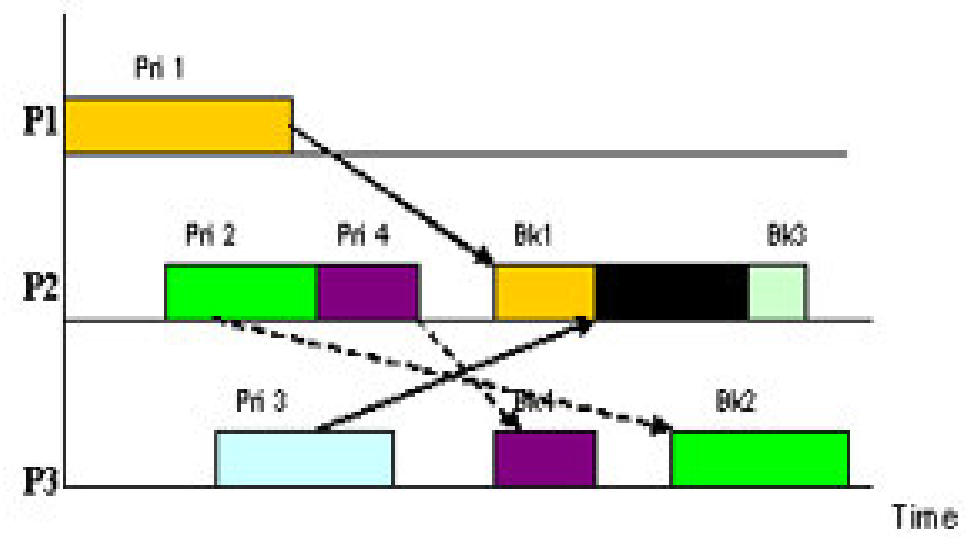

Fig. 3.1. View of the schedule.

Bk1 and Bk3 have been overloaded. As explained earlier, we can overload only those processes whose primaries are running on different processors. The arrows point from the primaries to their backup copies. Bk1 and Bk3 have been overloaded, i.e., they are scheduled to run at the same time, as discussed earlier, trying to perform backup overloading as far as possible
when there is a chance to do it.

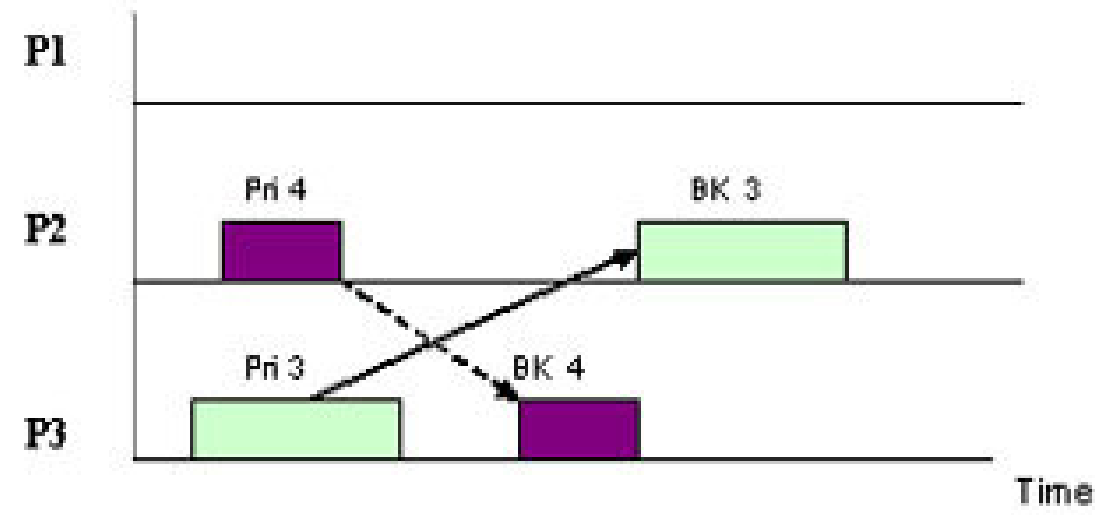

Fig. 3.2. View of the schedule.

Pri 1 and Pri 2 have finished execution based on EDF algorithm. So Bk1 and Bk2 are removed automatically.
Figure 3.2 shows how scheduling proceeds when primary1 and Primary 2 finish executing. According to the EDF schedule Pri 1 and Pri 2 have earliest deadlines and are executed first. Since there is no more need for backup 1 and 2, they are de-allocated from the processors in order to use CPU efficiently. This increases CPU utilization.

Figure 3.3 shows the schedule after 2 new processes (5 and 6) arrive. Again overloading is accomplished by scheduling Bk3 and Bk5 on the same time slot. Care is taken while scheduling not to overload 2 backups whose primary copies are on the same processor. Here primary 5 and 3 are on different processors 1 and 3 respectively.

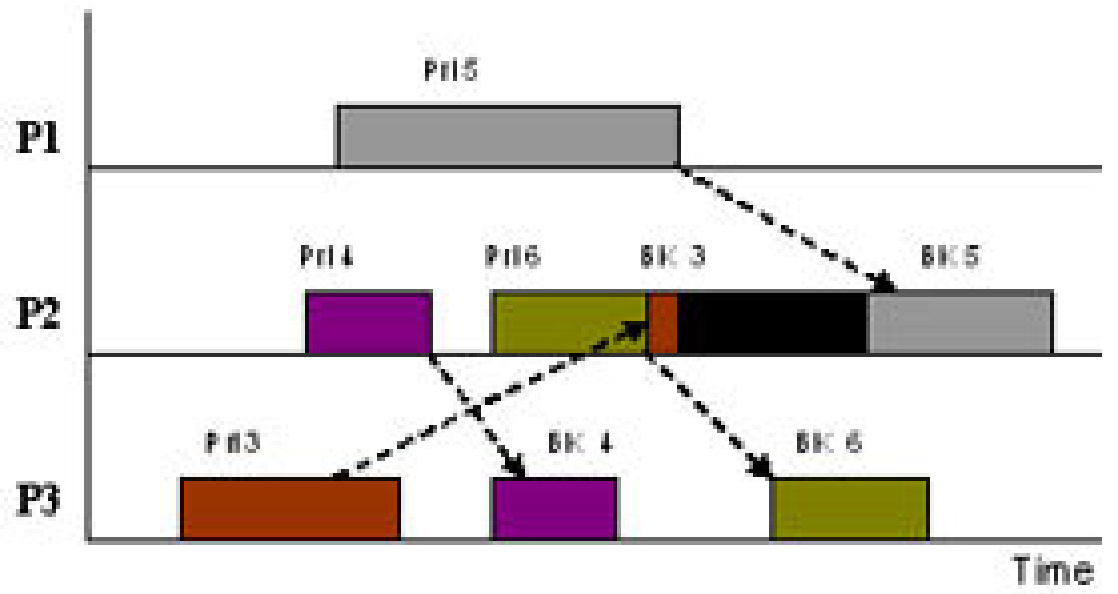

Fig. 3.3 View of the schedule. Backup copies of 5 and 6 are scheduled on P2 and P3. Again care is taken to see that backups of 4 and 6 are not overloaded.

## 4. CONCLUSION

The advantage of the algorithm is that the tasks considered are dynamic and aperiodic. The algorithm is simple and easy to implement. It also increases utilization speed and efficiency of scheduling. It can also be concluded that appropriate use of redundancy is important since too much redundancy increases reliability but potentially decreases the schedulability. Too little redundancy decreases reliability but increases schedulability.Also, designing, managing redundancy incurs additional cost, time, and memory and power consumption. Thus this algorithm can be efficiently used for fault tolerance in case where multiprocessors are used to run real-time applications.